\documentclass{aa}
\usepackage{txfonts}
\usepackage{graphicx}
\usepackage{txfonts}
\usepackage{natbib}
\usepackage{ulem}
\usepackage{booktabs}

\usepackage[colorlinks=true, linkcolor=blue, urlcolor=blue, citecolor=blue]{hyperref}

\usepackage{orcidlink}

\def\ll_lsun{log$({L/\rm L_{\odot}})$~}  
  
\def\masa_msun{$M/ \rm M_{\odot}$~}  
  
\def\m_mstar{$M/M_{*}$~}

\begin{document}

\title{Extreme mass loss during common envelope evolution: the origin of the double low-mass
  white dwarf system J2102--4145\thanks{The cooling sequences are publicly available at \protect\url{http://evolgroup.fcaglp.unlp.edu.ar}.}}

\author{
  Leandro G. Althaus\orcidlink{0000-0003-2771-7805}\inst{1}
\and
  Alejandro H. Córsico\orcidlink{0000-0002-0006-9900}\inst{2}
\and
  Mónica Zorotovic\orcidlink{0000-0002-4526-0469}\inst{3}
\and
  Maja V\v{u}\v{c}kovi\v{c}\orcidlink{0000-0001-8339-6423}\inst{3}
\and
  Alberto Rebassa-Mansergas\orcidlink{0000-0002-6153-7173}\inst{1,4}
\and
  Santiago Torres\orcidlink{0000-0001-5777-5251}\inst{1,4}
}

\institute{
Departament de Física, Universitat Politècnica de Catalunya,
c/ Esteve Terrades 5, 08860 Castelldefels, Spain
\and
Grupo de Evolución Estelar y Pulsaciones, Facultad de Ciencias Astronómicas y Geofísicas,
Universidad Nacional de La Plata, CONICET-IALP, Paseo del Bosque s/n,
1900 La Plata, Argentina
\\ \email{althaus@fcaglp.unlp.edu.ar}
\and
Instituto de Física y Astronomía, Universidad de Valparaíso,
Av. Gran Bretaña 1111, 5030 Casilla, Valparaíso, Chile
\and
Institut d’Estudis Espacials de Catalunya (IEEC),
C/ Esteve Terradas 1, Edifici RDIT, 08860 Castelldefels, Spain
}

\date{Received}  

\abstract
{Eclipsing close double white dwarf (WD) systems, by providing precise radii and masses, 
offer a unique opportunity to directly constrain hydrogen-envelope ($M_{\rm H}$) retention 
and test common-envelope (CE) evolution in low-mass stars.} 
{We analyze J2102--4145, an eclipsing binary composed of two low-mass He-core WDs in a 2.4-hour orbit.} 
{By comparing the observed radii and effective temperatures with updated CE and stable Roche-lobe overflow
(SRLOF) models, 
we confirm that both stars are He-core WDs.} 
{The primary ($0.375\,M_\odot$) is consistent with SRLOF models that retain thick H envelopes and sustain residual burning, 
while the secondary ($0.314\,M_\odot$) can only be reproduced by CE models with nearly complete envelope removal. 
The cooling ages---$\sim$220\,Myr for the secondary and $\sim$260–510\,Myr for the primary, depending on the residual nuclear contribution--- 
support a formation sequence in which the primary formed first via SRLOF, followed by a CE phase producing the compact secondary. 
Energy-budget reconstruction of the CE yields progenitor and orbital parameters consistent with this picture.} 
{The secondary’s unusually small radius requires an extremely thin H envelope, $M_{\rm H}\lesssim10^{-7}\,M_\odot$, 
well below the predictions of standard bifurcation criteria. 
J2102--4145 thus provides one of the strongest observational constraints on post-CE $M_{\rm H}$ in low-mass WDs 
and a benchmark challenge to current prescriptions of envelope ejection.}
\keywords{stars: evolution -- stars: interiors --
  stars: white dwarfs -- binaries: close -- stars: low-mass --
  methods: numerical} \titlerunning{CE Evolution and Cooling of He
  WDs} \maketitle
\authorrunning{Althaus et al.}  


\section{Introduction}
\label{intro}

White dwarfs (WDs) are the most common endpoints of low- and intermediate-mass stars
\citep{2010A&ARv..18..471A,2015ApJ...810...34W}. 
The WD mass distribution peaks near $0.6\,M_\odot$, 
but it also shows a low-mass tail ($\lesssim 0.45\,M_\odot$),
which likely results from enhanced red-giant branch (RGB) mass loss—most commonly driven by binary interaction—
before core helium (He) ignition, producing He-core rather than carbon–oxygen (C/O) cores
\citep{2013osp..book.....C,2013A&A...557A..19A,2016A&A...595A..35I}. 
Above $\sim0.32\,M_\odot$, some objects may be hybrid He--C/O WDs \citep{Zenati2019}.

Forming low-mass WDs within a Hubble time requires binary interaction, as single-star
evolution is too slow. Most such WDs are indeed found in binaries
\citep{1995MNRAS.275..828M,alberto2011,2020ApJ...889...49B}.
Two channels dominate the formation of He-core WDs: stable Roche-lobe overflow (SRLOF) and
common-envelope (CE) evolution. Population studies indicate that extremely low-mass (ELM) WDs with
$M\!\lesssim\!0.22\,M_{\sun}$ form predominantly via the SRLOF channel, whereas those with
$M\!\gtrsim\!0.22\,M_{\sun}$ are more likely the outcome of a CE episode
\citep{2019ApJ...871..148L,2020ApJ...889...49B}. 
Because the mass-loss histories differ, the post-formation structures and cooling behaviors
of low-mass WDs are observably distinct. A key quantity is the residual hydrogen (H) mass,
$M_{\rm H}$, defined here as the total H mass above the He core. In general, SRLOF leaves WDs
with relatively large values of $M_{\rm H}$, whereas CE evolution leads to
much more efficient envelope stripping and hence significantly smaller residual $M_{\rm H}$.
Larger $M_{\rm H}$ can sustain residual H burning and inflate the stellar radius, thereby
modifying the inferred mass and cooling age at fixed $T_{\rm eff}$
\citep{2013A&A...557A..19A,2016A&A...595A..35I}.

Eclipsing low-mass WD binaries provide precise, geometrically constrained masses and radii and
therefore enable stringent tests of structure, evolution, and formation scenarios
\citep{2023MNRAS.521.1880B}. In particular, J2102$-$4145 is a double-lined, eclipsing binary
of two low-mass WDs with accurately measured $M$, $R$, and $T_{\rm eff}$
\citep{2023ApJ...950..141K,amaral2024}. Here we confront these measurements with SRLOF models
\citep{2013A&A...557A..19A} and with CE sequences that span a broad range of $M_{\rm H}$, including
a recently published grid tailored to post-CE remnants \citep{althausCE}. We find that the
secondary (less massive) WD’s observed compactness at its $T_{\rm eff}$ is reproduced only
by models with an extremely thin  H envelope. Accordingly, we report a robust upper limit on $M_{\rm H}$.
For the primary (more massive) WD, we
reassess SRLOF models and quantify how residual H burning affects the inferred cooling age. Taken
together, these results support an evolutionary pathway in which the primary formed through SRLOF
and the secondary through a subsequent CE episode.

Sect.~\ref{obs} describes the observations and the adopted stellar parameters. 
Sect.~\ref{models} introduces the evolutionary models: Sect.~\ref{ce} presents the CE sequences 
and bifurcation criteria, while Sect.~\ref{srlof} describes the SRLOF tracks. 
Sect.~\ref{r_teff_cool} presents the radius–temperature comparison for J2102$-$4145 and the inferred 
envelope properties, including the possible effects of  rotation and tides. 
In this section it is also discussesed the corresponding cooling ages. 
Sect.~\ref{scenario} outlines the evolutionary pathway linking SRLOF for the primary and CE for the secondary. 
In Sect.~\ref{energy} we examine the relation between the inferred  $M_{\rm H}$ values and the CE 
efficiency, and discuss the implications for envelope stripping in low-mass 
giant progenitors.
Finally, Sect.~\ref{conclusions} summarizes our conclusions.

\section{Observations and adopted parameters}\label{obs}

J2102$-$4145 is a double-lined, eclipsing binary of two low-mass WDs. We adopt the
published orbital and stellar parameters from the joint modeling of multiband light
curves and radial velocities presented by \citet{amaral2024}. The orbital period is
$P=0.1002087525(10)\,{\rm d}$ ($\simeq 2.405$\,h) and the orbital inclination is
$i=88.693^{+0.006}_{-0.005}$ deg (Table~3 of \citealt{amaral2024}).
Eclipses provide the inclination and tight geometric constraints on the fractional
radii, while the RV semi-amplitudes yield dynamical masses through Kepler’s law.
The eclipse morphology (ingress/egress durations) sets $R/a$, and the relative
depths constrain the surface-brightness ratio; combining these with the RV solution
gives $M$ and $R$ for each component with high precision. Spectroscopic fits provide
$T_{\rm eff}$ values; throughout this work we use the $1\sigma$ uncertainties reported
by \citet{amaral2024} and do not re-analyze the photometric or spectroscopic data.

Table~\ref{tab:obs} lists the masses, radii, and $T_{\rm eff}$ adopted here for the
primary and secondary. These observables constitute the data vector against which we
confront SRLOF and CE evolutionary models in Sect.~\ref{r_teff_cool}.
When inferring cooling ages, we evaluate the model age at the measured $T_{\rm eff}$
of each component and propagate the observational uncertainty in $T_{\rm eff}$ along
the corresponding track at fixed $M$ and, for CE models, fixed $M_{\rm H}$.
In the $R$--$T_{\rm eff}$ comparison we use the quoted
$1\sigma$ error bars on both quantities.

\begin{table}
\centering
\caption{Observed parameters for J2102--4145 adopted in this work.}
\label{tab:obs}
\begin{tabular}{lccc}
\hline\hline
Component & $M\ (M_\odot)$ & $R\ (R_\odot)$ & $T_{\rm eff}\ ({\rm K})$ \\
\hline
Primary   & $0.375\pm0.003$ & $0.0211\pm0.0002$ & $13688^{+65}_{-72}$ \\
Secondary & $0.314\pm0.003$ & $0.0203^{+0.0002}_{-0.0003}$ & $12952^{+53}_{-66}$ \\
\hline
\end{tabular}
\\[1ex]
\footnotesize{Values from \citet{amaral2024}; uncertainties are $1\sigma$.}
\end{table}

\begin{figure}
        \centering
        \includegraphics[width=1.\columnwidth]{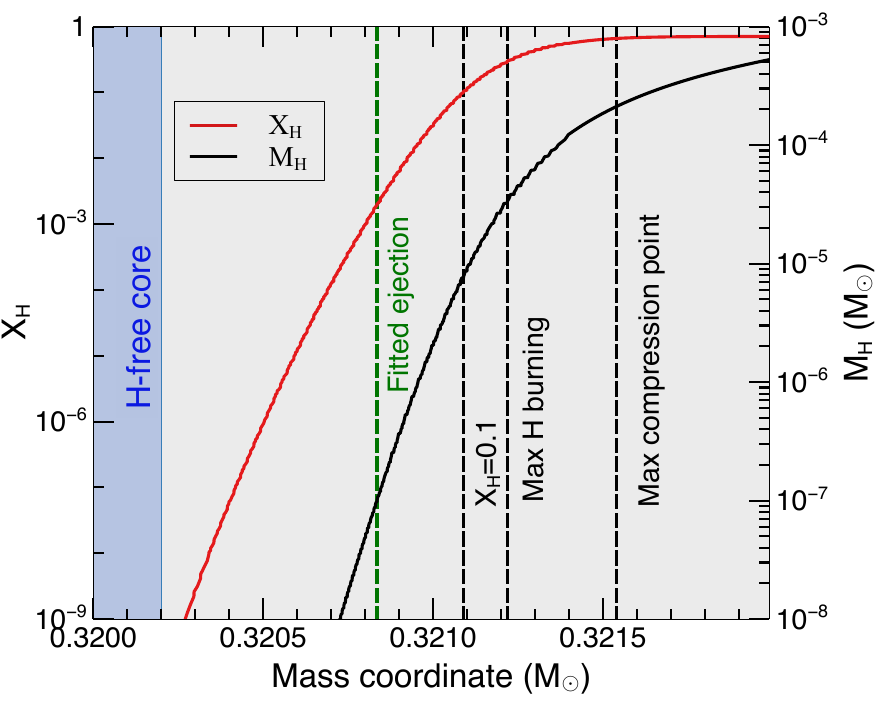}
\caption{
Internal structure of a $1.5\,M_\odot$ pre-CE RGB star at the epoch when the H-free core is
$0.3202\,M_\odot$ (plotted versus the Lagrangian mass coordinate $m_r$). Shown are the H abundance
by mass, $X_{\rm H}(m_r)$, and the cumulative H mass interior to $m_r$,
$M_{\rm H}(<m_r)$, which increases outward. Vertical dashed lines mark the BP criteria
($X_{\rm H}=0.1$, peak nuclear energy generation, and maximum compression). The dashed green line
marks $M_{\rm H}=1.0\times10^{-7}\,M_\odot$, the H content inferred for the secondary of
J2102--4145 from the observed $(R, T_{\rm eff})$. The plot shows only the innermost envelope
layers of the  RGB progenitor, which set the residual H content.
}
        \label{bifurcation}
\end{figure}

\section{Evolutionary models}\label{models}

In this section we outline the evolutionary inputs used to interpret J2102$-$4145, focusing on the
single parameter that drives the differences between channels—$M_{\rm H}$,
which controls radius inflation and residual burning. Sect.~\ref{ce} summarizes the CE construction
and the compression-point guided stripping adopted for the secondary; Sect.~\ref{srlof} recalls the
SRLOF grids used for the primary and how we bracket the effect of residual burning.

\subsection{CE sequences: setup, bifurcation criteria, and thin-envelope behaviour}\label{ce}

We adopt the CE grid tailored to post-CE remnants by \citet{althausCE}. A solar-metallicity RGB
progenitor is evolved until the H-free core reaches the target mass; 
the envelope is then removed abruptly down to a prescribed Lagrangian mass coordinate $m_r$
(i.e., the enclosed mass measured from the stellar center), which sets the remaining $M_{\rm H}$.
The stripped
model is relaxed at fixed total mass and evolved with the La Plata stellar evolution code {\tt LPCODE}
\citep{2005A&A...435..631A,2013A&A...555A..96S,2015A&A...576A...9A,2016A&A...588A..25M,2022A&A...663A.167A}
onto the cooling track. Because mass stripping proceeds much faster than the thermal
timescale, the remnant contracts to the hot pre-WD phase on a timescale much shorter than its thermal
time; for the thin-$M_{\rm H}$ regime of interest here, typical adjustment times from ejection to the
pre-WD maximum $T_{\rm eff}$ are below about $0.01$ Myr.
Shortly after stripping, the outer layers are He-enriched; as the star settles on the WD branch,
gravitational settling builds a pure-H surface above a He-rich tail, with measurable consequences
for $R$, $\log g$, and cooling rates \citep{althausCE}.

After a dynamical CE, some H remains bound to the core rather than being fully expelled
\citep{lombardi2006,ivanova2011}; this final $M_{\rm H}$ is crucial for the subsequent evolution
of the He WD. A major modelling uncertainty is the location of the bifurcation point (BP), the
mass coordinate that separates the retained core from the ejected envelope. In practice, the BP
lies within the H-rich layers, between the H-free core and the base of the convective envelope.
Several physically motivated criteria have been proposed to identify this boundary, including
features in the entropy profile, breaks in the binding-energy distribution, and diagnostics tied
to the structure of the H-burning shell
\citep{han1994,tauridewi2001,ivanova2013review,2016A&A...596A..58K,2022MNRAS.511.2326V,chenreview}.
In this work, we focus on three commonly adopted BP diagnostics that are directly relevant for our
CE models: the layer where the H abundance drops to $X_{\rm H}=0.1$, the location of peak nuclear
energy generation in the H-burning shell, and the maximum compression point $m_{\rm cp}$, defined
by the local maximum of $P/\rho$ inside the H-burning shell prior to CE \citep{ivanova2011}.
Remnants stripped to or below $m_{\rm cp}$ relax by contracting; if the remnant is left above
$m_{\rm cp}$ it tends to re-expand on a local thermal timescale and shed additional mass until
stabilizing near $m_{\rm cp}$. Thus, $m_{\rm cp}$ provides a physically meaningful core--envelope
boundary for CE outcomes and naturally leads to comparatively thin residual H contents.

For reference, Fig.~\ref{bifurcation} shows the internal structure of a pre-CE RGB progenitor with
$1.5\,M_\odot$ at the epoch when the H-free core reaches $0.3202\,M_\odot$. We plot the H abundance
profile $X_{\rm H}(m_r)$ and the cumulative H mass interior to the Lagrangian mass coordinate
$m_r$, $M_{\rm H}(<m_r)$, which increases outward through the envelope. Vertical dashed lines mark
the three BP diagnostics discussed above ($X_{\rm H}=0.1$, peak nuclear energy generation, and
maximum compression). The dashed green line indicates $M_{\rm H}=1.0\times10^{-7}\,M_\odot$, the
H content required to reproduce the observed radius of the secondary at its measured
$T_{\rm eff}$ (Sect.~\ref{r_teff_cool}). Although the figure corresponds to a $1.5\,M_\odot$
progenitor, it serves as an illustrative snapshot of how the remaining H content maps to the
stripping depth within the H-rich layers. This qualitative correspondence applies directly to
the CE sequences employed in this work.

From the published set we employ the CE tracks at $M=0.3208\,M_\odot$ with $M_{\rm H}=6.6\times 10^{-6}\,M_\odot$
and at $M=0.363\,M_\odot$ with $M_{\rm H}=5\times 10^{-6}\,M_\odot$, and we computed two additional thin-envelope
sequences at the same $M=0.3208\,M_\odot$ with $M_{\rm H}=10^{-6}$ and $10^{-7}\,M_\odot$ to probe the regime
relevant to the secondary of J2102$-$4145. These are the CE tracks used in the $R$–$T_{\rm eff}$ and $T_{\rm eff}$–age
comparisons (Sect.~\ref{r_teff_cool}).

\subsection{SRLOF sequences}\label{srlof}

For the primary we use SRLOF He-core WD sequences from \citet{2013A&A...557A..19A}. In contrast to
our CE remnants with very thin envelopes, SRLOF histories leave thicker H layers that sustain
residual H burning during much of the cooling phase. This tends to inflate the radius at a given
$T_{\rm eff}$ and to lengthen the cooling timescales at intermediate temperatures. In the mass range
relevant here, the SRLOF track that best matches the primary has an envelope of order
$M_{\rm H}\sim {\rm few}\times 10^{-4}\,M_\odot$, i.e., orders of magnitude larger than in the thin–$M_{\rm H}$
CE regime considered for the secondary.
In this work we therefore consider a representative SRLOF track near the primary’s mass (we use
$M \simeq 0.363\,M_\odot$ as our reference) and bracket the impact of residual burning with two
configurations: a case with $L_{\rm nuc}/L\simeq 0.25$ (as in \citealt{2013A&A...557A..19A}) and a case with
enhanced residual burning, $L_{\rm nuc}/L\simeq0.70$, 
to illustrate how the inferred cooling age depends on the nuclear contribution. 
These SRLOF models are then used in Sect.~\ref{r_teff_cool} to interpret the primary.
Over the $M$–$T_{\rm eff}$ range of interest, the structural parameters and cooling ages predicted by
these SRLOF sequences are consistent—within typical modelling differences—with independent calculations
(e.g., \citealt{2016A&A...595A..35I}). This agreement provides a useful cross-check on our
radius–temperature comparison and on the cooling-age estimates for the primary.

\begin{figure}
        \centering
        \includegraphics[width=1.\columnwidth]{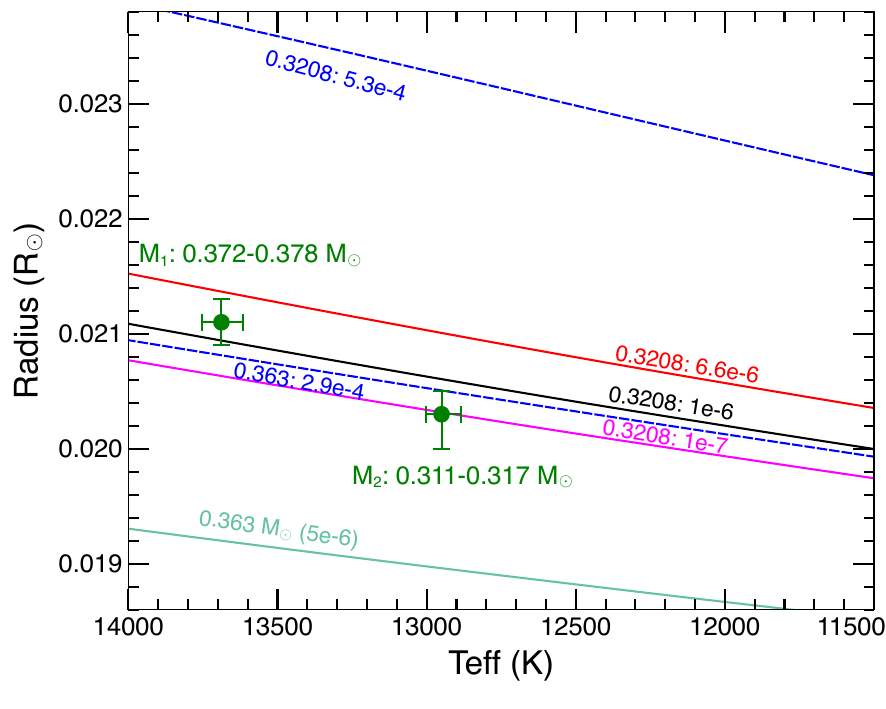}
\caption{Stellar radius ($R_\odot$) versus $T_{\rm eff}$ for He-core WD sequences with different $M_{\rm H}$. 
Shown are CE models at $M=0.3208\,M_\odot$ ($M_{\rm H}=6.6\times10^{-6}$, $10^{-6}$, and $10^{-7}\,M_\odot$) 
and at $M=0.363\,M_\odot$ ($M_{\rm H}=5\times10^{-6}\,M_\odot$) \citep{althausCE}, together with SRLOF tracks 
from \citet{2013A&A...557A..19A} (blue dashed). The J2102--4145 components are shown with $1\sigma$ errors in 
$R$ and $T_{\rm eff}$, including the observed mass ranges of the primary ($M_1$) and secondary ($M_2$) 
\citep{amaral2024}. The curve annotations in the plot indicate the stellar mass and H envelope mass, in the format 
$M$:$M_{\rm H}$ (both in $M_\odot$), for each evolutionary sequence. The primary matches an SRLOF model 
with $M_{\rm H}\sim3\times10^{-4}\,M_\odot$, while the secondary requires $M_{\rm H}\lesssim10^{-7}\,M_\odot$ 
(see text).
}
\label{rteff}
\end{figure}

\section{Radius--temperature comparison and cooling ages}\label{r_teff_cool}

In Fig.~\ref{rteff} we show CE tracks at $M=0.3208\,M_\odot$ with
$M_{\rm H}=6.6\times10^{-6}\,M_\odot$ \citep{althausCE} and the two additional thin-envelope
CE sequences computed here at the same mass with $M_{\rm H}=10^{-6}$ and $10^{-7}\,M_\odot$.
We also include the published CE sequence at $M=0.363\,M_\odot$ with
$M_{\rm H}=5\times10^{-6}\,M_\odot$ \citep{althausCE}. For comparison, we plot SRLOF models from
\citet{2013A&A...557A..19A} that retain $M_{\rm H}\sim(3$–$5)\times10^{-4}\,M_\odot$ and sustain
residual H burning (shown as blue dashed lines). The locations of the primary and secondary
(Table~\ref{tab:obs}; \citealt{amaral2024}) are overplotted with their $1\sigma$ intervals in
$R$ and $T_{\rm eff}$.

The comparison reveals a clear structural dichotomy. Despite being less massive, the secondary
($M_2$) has a slightly smaller radius than the primary ($M_1$), contrary to
the mass--radius relation expected for similar compositions. Since both stars have comparable
$T_{\rm eff}$ (within $\sim 800$\,K), this inversion cannot be ascribed to thermal differences and
instead points to distinct envelope structures and evolutionary histories.

The secondary cannot be reproduced by the SRLOF track at its measured mass. In particular, the
$0.3208\,M_\odot$ SRLOF sequence predicts radii
systematically larger than $R_2$ at the observed $T_{\rm eff}$. Matching $R_2$ along this track would require
moving to much cooler $T_{\rm eff}$, inconsistent with the data. We therefore focus on CE models. At the
secondary’s $T_{\rm eff}$, CE models predict a monotonic decrease of $R$ with decreasing $M_{\rm H}$. The
$M=0.3208\,M_\odot$ sequences with $M_{\rm H}=6.6\times10^{-6}$ and $10^{-6}\,M_\odot$ overpredict the measured
$R_2$ (the latter only marginally within the $1\sigma$ band), whereas the $10^{-7}\,M_\odot$ track matches the
observed radius. This is a conservative value: at fixed $T_{\rm eff}$ a slightly lower mass implies a larger
radius, so using the available $0.3208\,M_\odot$ tracks instead of a track closer to the dynamical mass mildly
underestimates the model radius and would require an even thinner envelope to match $R_2$. If $M=0.317\,M_\odot$,
the $M_{\rm H}=10^{-7}\,M_\odot$ track would pass through the upper bound of $R_2$ (i.e., within $+1\sigma$), so
our adopted $M_{\rm H}\lesssim10^{-7}\,M_\odot$ remains a conservative, observationally anchored limit.

The extremely low $M_{\rm H}$ implied by the secondary raises the question of whether standard
BP prescriptions in CE evolution can produce such remnants. For the progenitor
structure shown in Fig.~\ref{bifurcation} (a $1.5\,M_\odot$ RGB star at
$M_{\rm core}=0.3202\,M_\odot$), the $M_{\rm H}$ values associated with the maximum-compression
point ($m_{\rm cp}$) and with the $X_{\rm H}=0.1$ proxy are both far above what is required by $R_2$.
In fact, $M_{\rm H}\simeq 1.0\times10^{-7}\,M_\odot$ is roughly three orders of magnitude lower than
the value at $m_{\rm cp}$ and about sixty times smaller than the $X_{\rm H}=0.1$ estimate (see
Fig.~\ref{bifurcation}). This contrast indicates that envelope removal in J2102--4145 was
substantially more efficient than predicted by standard BP criteria.

\begin{figure}
        \centering
        \includegraphics[width=1.\columnwidth]{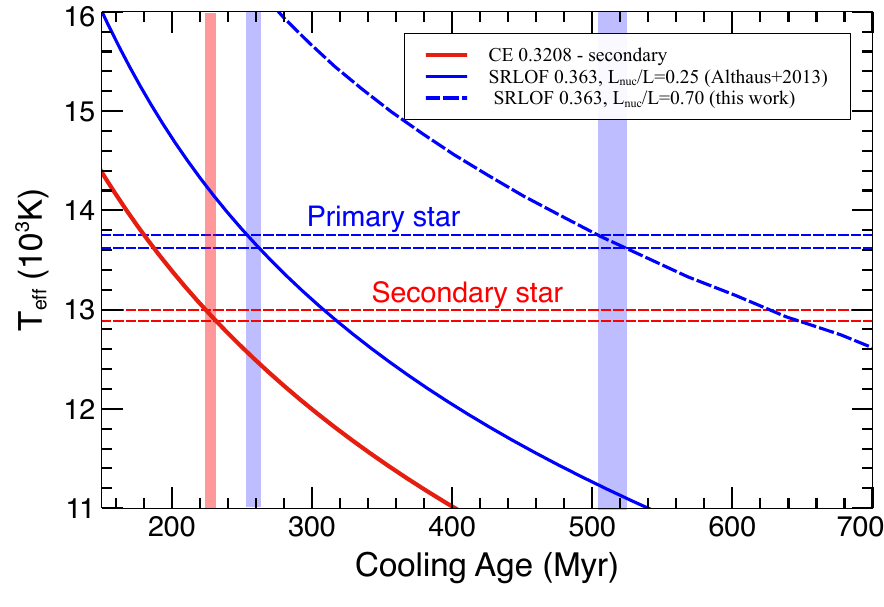}
\caption{
Evolution of $T_{\rm eff}$ since the end of mass loss for selected He-core WD models.
The red curve shows the CE track ($M=0.3208\,M_\odot$, $M_{\rm H}=10^{-7}\,M_\odot$) used for the
secondary. Solid blue is the post-SRLOF model for the primary from \citet{2013A&A...557A..19A}
with residual burning $L_{\rm nuc}/L\simeq0.25$, and dashed blue is a variant computed here with
increased residual burning ($L_{\rm nuc}/L\simeq0.70$) to illustrate its impact on the cooling.
Colored bars indicate the cooling ages inferred at the measured
$T_{\rm eff}$ of each component, with their widths obtained by propagating the observational
uncertainty in $T_{\rm eff}$ along the corresponding track (at fixed $M$ and $M_{\rm H}$).
}
        \label{teffage}
\end{figure}

The primary is well reproduced by SRLOF models that retain comparatively thick H envelopes
($M_{\rm H}\sim 10^{-4}\,M_\odot$), which inflate the radius and sustain
residual H burning. In contrast, the secondary—already shown to require an extremely thin
envelope of $M_{\rm H}\simeq 1\times10^{-7}\,M_\odot$—can only be explained by nearly complete
envelope removal during CE. The juxtaposition of these two cases strongly points to distinct
evolutionary channels.
 \citet{amaral2024} also
found the primary consistent with substantial H retention using independent SRLOF grids
\citep{2016A&A...595A..35I}, but proposed that the secondary formed first via SRLOF and the primary
later through CE. Our structural analysis supports the reverse order: the primary likely formed via
SRLOF, retaining a relatively massive H layer, and the secondary through a subsequent CE episode
that stripped almost all of its envelope. We return to this formation sequence in Sect.~\ref{scenario}.

Element diffusion is an unavoidable process during WD cooling, as
gravitational settling reshapes the outer layers and generally
increases the stellar radius at a given $T_{\rm eff}$. To assess its
impact, we computed additional sequences in which the effects of
element diffusion were artificially suppressed. In this case, matching
the observed radius of the secondary requires a thicker H envelope,
$M_{\rm H}\sim 6\times10^{-6}\,M_\odot$—still a very thin layer
compared with SRLOF remnants. This value is close to that predicted by
the $X_{\rm H}=0.1$ bifurcation criterion and, while smaller than the
estimate obtained from the compression point by a factor of a
few–tens, it is not orders of magnitude different.

Suppressing diffusion, however, requires invoking an additional
mechanism capable of competing with the settling timescale. A plausible
candidate is rotationally induced mixing. One way to estimate the
associated transport is through the Eddington–Sweet circulation
velocity, given by
\begin{equation}
v_{\rm ES} \;\sim\; \frac{\Omega^{2}\,R^{5}\,L}{G^{2}\,M^{3}} \, ,
\end{equation}
where $\Omega$ is the stellar angular velocity. If the secondary is
synchronized with the orbit, $\Omega$ corresponds to the orbital
frequency ($P=0.07$\,d), which provides an upper limit on the rotation
rate and thus on $v_{\rm ES}$. For luminous post-CE models, this yields
order-of-magnitude flow speeds in the H-rich envelope that can exceed
the diffusion velocities by about one to two orders of magnitude.
Because $v_{\rm ES}\propto L$, the efficiency of this process decreases
as the star cools, and in practice composition gradients and turbulence
may further reduce the effective mixing. Nevertheless, the comparison
suggests that rotationally driven transport could, in some regimes,
compete with gravitational settling and modify the  envelope
structure.

The purpose of invoking rotation here is mainly to highlight a
mechanism that could inhibit diffusion and thereby relax the
constraint on the required $M_{\rm H}$. Rotation may also influence the
stellar structure more directly—for instance, centrifugal support would
tend to increase the stellar radius, which in turn would require
thinner envelopes to match $R_2$. Thus, envelope and core rotation
could act in opposite directions regarding the inferred $M_{\rm H}$,
but in all cases the required H layers remain extremely thin. A
detailed assessment of these effects lies beyond the scope of this
work, yet they underline the sensitivity of our inferences to
rotational physics.

By contrast, tidal heating—another possible perturbation—remains
negligible for orbital periods $\gtrsim 1$\,h \citep[e.g.,][]{fuller2013},
and is unlikely to affect the present system given its 2.4\,h nearly
circular orbit \citep{amaral2024}.

Fig.~\ref{teffage} illustrates the $T_{\rm eff}$ evolution of the
reference models. For the secondary, we adopt the CE track with
$M=0.3208\,M_\odot$ and $M_{\rm H}=10^{-7}\,M_\odot$, which reaches the
observed parameters after $\sim$220\,Myr of cooling. For the primary,
we rely on the representative SRLOF sequences introduced in
Sect.~\ref{srlof}. The case with $L_{\rm nuc}/L\simeq0.25$
(solid blue line; \citealt{2013A&A...557A..19A}) reproduces the
observed $T_{\rm eff}$ at an age of $\sim$260\,Myr. A variant
sequence with enhanced burning, $L_{\rm nuc}/L\simeq0.70$ (dashed blue
line), illustrates the sensitivity of the result: at the same
$T_{\rm eff}$, the inferred cooling age can increase to
$\sim$510\,Myr. This should not be regarded as a firm upper limit, but
rather as an indication of how strongly the age estimate depends on
the contribution of residual nuclear burning. The $\sim$240\,Myr age
inferred from an independent $0.344\,M_\odot$ sequence by
\citet{2016A&A...595A..35I}, with a nuclear contribution of
$\sim$30\%, further supports the consistency of our adopted SRLOF
models. 

The cooling time of the primary thus depends critically on the extent
of residual burning, which in turn is set by the final envelope mass.
As shown by \citet{althaus2001}, this depends on the number of CNO
flashes experienced before entering the cooling track, a quantity that
increases when element diffusion is included. The efficiency of
diffusion in the outer H tail—where it dominates—may itself be
modified by degeneracy, turbulence, or rotation-induced mixing,
highlighting that the predicted residual nuclear burning, and hence the
cooling age, remain sensitive to these physical uncertainties.

Because post-CE models cool faster than SRLOF sequences,
and given the observed $T_{\rm eff}$ values,
the system configuration strongly favors the primary having formed first via SRLOF,
followed by a CE episode that produced the compact secondary.
This contrasts with the scenario of \citet{amaral2024}, where the more massive WD forms last via CE.

\section{Formation history and evolutionary pathway of J2102--4145}
\label{scenario}

The measurable age difference between the two WDs in J2102--4145
provides strong constraints on its evolution. Our models indicate that
the primary—an SRLOF-formed $0.375\,M_\odot$ He-core WD—reaches its
current $T_{\rm eff}$ after $\sim$260\,Myr for a case with
$L_{\rm nuc}/L\simeq0.25$, and after $\sim$510\,Myr in an illustrative
model with enhanced residual burning ($L_{\rm nuc}/L\simeq0.70$); even
larger nuclear contributions would lengthen the cooling time further.
The secondary—formed through a CE event—matches its properties in
$\sim$220\,Myr. Thus, the delay between the SRLOF episode and the CE
event is of order a few $10^8$\,yr; within our bracketing models it lies
in the range $\sim$40–290\,Myr.

A further constraint comes from the primary’s radius.
Although its mass does not exclude a C/O core \citep{Zenati2019},
such models predict a much smaller radius due to stronger Coulomb interactions in the degenerate plasma,
even with the largest $M_{\rm H}$ allowed after SRLOF.
The observed radius is instead reproduced by a He-core model with residual burning,
confirming that core He ignition was avoided.
The secondary, with $0.314\,M_\odot$,  lies below the helium ignition threshold,
making a C/O core implausible.

Avoiding core He ignition requires a primary progenitor mass below $\sim1.8\,M_\odot$
\citep[e.g.,][]{hanetal2002,Arancibiarojasetal24}.
A higher mass ($\gtrsim 3.5$–$4\,M_\odot$) could work if the envelope were lost on the main sequence,
but mass transfer is then slow and unlikely to remove the envelope before the orbit widens,
detaching the donor. This would demand an initial period $\lesssim 1$–2\,d,
typical only of hierarchical triples \citep{Tokovinin2006}.

The stability of the first mass-transfer episode depends on the donor’s structure and mass ratio.
Beyond the Hertzsprung gap, partially convective envelopes tend to expand during mass loss,
favouring CE, but stable transfer is possible if $q \lesssim 1.5$–2.2
\citep[e.g.,][]{pavlovski2015}.
Earlier transfer, during the main sequence or subgiant phase, allows larger $q$ \citep{ge2024}.
Our scenario thus requires moderately constrained parameters but no extreme fine tuning.

A plausible pathway begins with SRLOF from the more massive progenitor,
forming the primary with a relatively large $M_{\rm H}$.
Mass transfer may increase the secondary’s mass, shortening its main-sequence lifetime
and allowing it to reach the RGB and undergo CE within $\sim$280\,Myr.
For example, a $1.8\,M_\odot$ star leaves the main sequence after $\sim$1600\,Myr,
while a $1.6\,M_\odot$ star does so after $\sim$2250\,Myr—too late for the required timing
unless mass accretion accelerates its evolution.
The retained fraction is uncertain: some studies support non-conservative transfer
\citep[e.g.,][]{hanetal2002,hanetal2003,vanRensbergen2008,vanRensbergen2011,Erdem2014,Marino2019},
others predict negligible accretion due to spin-up \citep[e.g.,][]{vos2020},
while up to 50\% retention is also possible \citep{lechien2025}.

\begin{figure}
        \centering
        \includegraphics[width=.8\columnwidth]{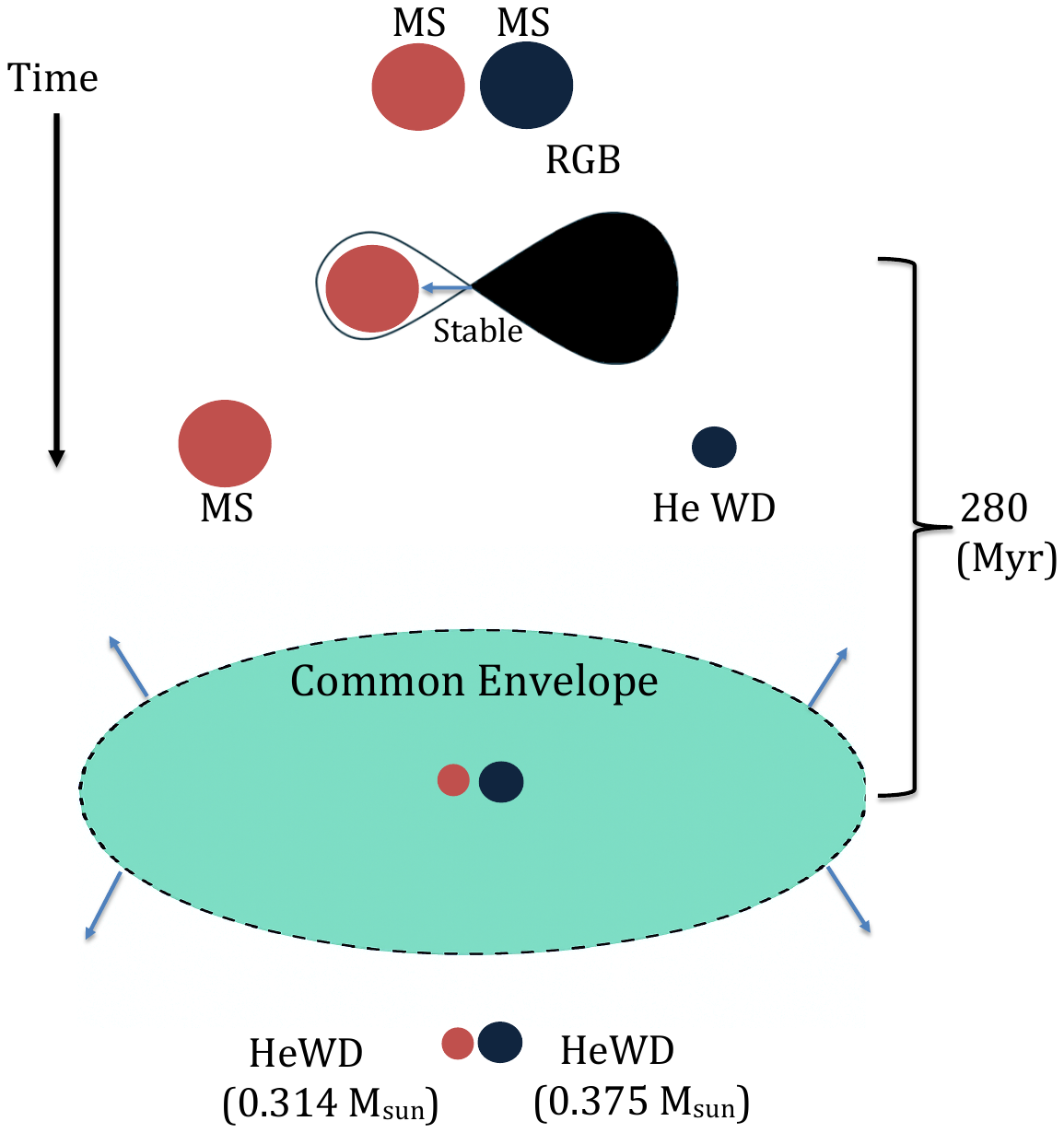}
\vspace{-40pt} 
\caption{Schematic diagram illustrating the proposed formation sequence of J2102--4145, including the
  two mass-transfer episodes. Stellar separations are not to scale.
}
\label{RL}
\end{figure}

An alternative explanation for the observed age difference is that the
system began as a near-equal-mass binary, consistent with the twin
peak in the mass-ratio distribution of main-sequence binaries,
particularly in short-period systems \citep[e.g.,][]{Halbwachs2003}.
With nearly equal masses, the first episode of stable mass transfer
(SRLOF) is expected to widen the orbit.  If the orbit widens, the
secondary would encounter a larger Roche lobe at the second
interaction and would therefore need to reach a larger stellar radius
to initiate mass transfer.  For RGB donors, a larger radius typically
implies a larger He-core mass, so this path would predict a more
massive resulting WD for the secondary—contrary to the observations.

The orbital response to mass transfer depends on how mass and angular
momentum are lost from the system \citep{VosVuckovic2017}.  Mass loss
from the vicinity of the donor or accretor generally leads to orbital
widening, whereas losses through a circumbinary structure can extract
angular momentum and cause orbital shrinkage.  While such effects are
discussed in the literature, they are not required to account for the
properties of J2102--4145.

The different masses of the two WDs are naturally explained by the
distinct timescales of the two interaction phases.  Stable Roche-lobe
overflow proceeds over tens of Myr, allowing the donor to continue its
RGB evolution while the orbit evolves gradually and the He core grows.
By contrast, the CE phase experienced by the secondary is dynamical
and strips the envelope on a short timescale, with little additional
core growth.  As a result, the secondary can reach Roche-lobe overflow
at its present orbital separation with a smaller He-core mass and end
as the less massive WD, without invoking any special orbital evolution
prior to the CE.

Applying the reconstruction method of \citet{2022MNRAS.513.3587Z} to the CE event,
we find that the progenitor of the less massive WD likely had a mass of $\sim$1.1–1.9\,$M_\odot$
and an orbital period of $\sim$20–60\,days when filling its Roche lobe.
These values match the observed masses and cooling ages, supporting the proposed scenario.
Fig.~\ref{RL} illustrates the pathway, including both mass-transfer phases.

In closing, we stress that the cooling ages imply that the CE episode must have occurred within
$\lesssim 200$--$300\,\mathrm{Myr}$ after the end of the first RLOF phase.
This requirement can be met if the secondary progenitor is already close to the
terminal-age main sequence when the primary detaches from RLOF, and/or if it is
modestly rejuvenated by accretion.
To illustrate the relevant evolutionary timescales, we note that in our
solar-metallicity LPCODE tracks the time elapsed from the end of central H burning
to the point where the He-core mass reaches $\simeq 0.31\,M_\odot$ is
$\sim 300\,\mathrm{Myr}$ for a $1.5\,M_\odot$ star, but decreases to
$\sim 130\,\mathrm{Myr}$ for a $1.8\,M_\odot$ star.
Therefore, once the secondary has left the main sequence, reaching the RGB core
mass required for the CE phase is compatible with the time window inferred from
the WD cooling ages.

\section{Energy budget of the common-envelope ejection}
\label{energy}

A quantitative description of the CE phase can be
formulated in terms of the energy formalism \citep{ivanova2013review},
in which the orbital energy released during the spiral-in process is
compared with the binding energy of the donor’s envelope. The ejection
efficiency is described by the parameter $\alpha_{\rm CE}$, defined as

\begin{equation}
\alpha_{\rm CE} \, \Delta E_{\rm orb} = E_{\rm bind},
\label{eq:alpha_ce}
\end{equation}
where \(\Delta E_{\rm orb}\) is the orbital energy released between
the initial (\(a_{\rm i}\)) and final (\(a_{\rm f}\)) separations, and \(E_{\rm
  bind}\) is the total binding energy of the envelope at the onset of
CE.

For the J2102--4145 system, the energy analysis presented here refers
to the CE episode that formed the secondary WD of  \(M_{2,{\rm
    WD}}=0.314\,M_\odot\). At that stage, the
primary was already a WD of mass \(M_{1,{\rm WD}}=0.375\,M_\odot\),
while the companion was the RGB progenitor of the present-day
secondary. In this section, we consider progenitor models for the
secondary with zero-age main sequence masses \(M_{\rm ZAMS,2}\)
between 1.0 and 1.8\,$M_\odot$.
The system currently has
an orbital period \(P=0.10020875\) d, corresponding to a separation
\(a_{\rm now}=0.80190\,R_\odot\), and a cooling age for the secondary
of \(\tau_{\rm cool,2}=220\) Myr. By backward integrating the
angular-momentum losses driven by gravitational radiation for a
circular orbit \citep{peters1964} over \(\tau_{\rm cool,2}\), we
derive a post-CE (``birth'') separation \(a_{\rm
  birth}=0.85418\,R_\odot\), which we adopt as the final orbital
separation of the CE episode (\(a_{\rm f}=a_{\rm birth}\)).

The orbital energy released during the spiral-in of the RGB
progenitor of the secondary around the pre-existing WD is
given by

\begin{equation}
\Delta E_{\rm orb} \;=\; 
\frac{G\,M_{1,{\rm WD}}\,M_{2,{\rm WD}}}{2\,a_{\rm f}}
\;-\;
\frac{G\,M_{1,{\rm WD}}\,M_{\rm ZAMS,2}}{2\,a_{\rm i}}\,,
\label{eq:deltaEorb}
\end{equation}
The initial separation \(a_{\rm i}\) is determined from the Roche-lobe overflow 
condition of the RGB donor at the onset of the CE. To estimate \(a_{\rm i}\), we 
use the photospheric radius of the star (\(R_{\rm phot}\)) as a function of the mass of its 
H-free core and identify the point where this mass is approximately equal 
to the final mass of the secondary WD, \(M_{2,{\rm WD}}\). The resulting 
\(R_{\rm phot}\) and corresponding \(a_{\rm i}\) values for 
\(M_{\rm ZAMS,2}=\{1.0,\,1.5,\,1.8\}\,M_\odot\) are listed in 
Table~\ref{tab:master_CE}. Because \(a_{\rm i} \gg a_{\rm f}\), the change in orbital 
energy is largely determined by the \(a_{\rm f}\) term in 
Eq.~(\ref{eq:deltaEorb}).

For the energy budget, the pre-CE boundary between the core and the ejected 
envelope is set by a mass coordinate \(m_{\rm cut}\) linked to the 
observed radius of the secondary WD. It is defined so that the total 
H mass enclosed within \(m_{\rm cut}\) in the pre-CE RGB model equals 
the H mass (\(M_{\rm H}^{\rm fit}\)) that, after envelope ejection, 
reproduces the WD radius (see Fig. \ref{bifurcation}),
\begin{equation}
\int_{0}^{m_{\rm cut}} X_{\rm H}(m)\,{\rm d}m \;=\; M_{\rm H}^{\rm fit},
\qquad M_{\rm H}^{\rm fit}=10^{-7}\,M_\odot.
\label{eq:mcut}
\end{equation}
The binding energy of the outer envelope (above \(m_{\rm cut}\)) is then computed as

\begin{equation}
E_{\rm bind}(m_{\rm cut}) \;=\;
\int_{m_{\rm cut}}^{M_{\rm ZAMS}}
\left[
\frac{G\,m}{r(m)} \;-\; u_{\rm int}(m)
\right]\,{\rm d}m,
\label{eq:Ebind}
\end{equation}

where \(u_{\rm int}\) is the specific internal energy from the OPAL 
equation of state, which includes gas, radiation, and ionization terms 
appropriate to the pre-CE structure. No enthalpy contribution 
\((+P/\rho)\) or recombination energy released during the CE expansion 
is added. Hence, 
our \(E_{\rm bind}\) represents a moderately conservative estimate: 
smaller than a purely gravitational value, but larger than those 
including enthalpy or  recombination energy liberated during 
the CE expansion.

Our values of $\alpha_{\rm CE}$, summarized in 
Table~\ref{tab:master_CE}, show the expected trend with progenitor mass: 
more massive RGB donors have more tightly bound envelopes 
(higher $E_{\rm bind}$) and thus require larger ejection efficiencies. 
The $1.0\,M_\odot$ progenitor yields the smallest $\alpha_{\rm CE}$, 
but it would evolve too slowly to reproduce the $\sim0.2$\,Gyr formation 
delay between the two WDs. In contrast, the $1.8\,M_\odot$ model requires 
$\alpha_{\rm CE}>1$, implying that orbital energy alone would be 
insufficient unless additional power sources such as jets, nuclear 
burning, or dynamically released recombination energy were available. 
The intermediate $1.5\,M_\odot$ progenitor yields 
$\alpha_{\rm CE}\simeq 1$, providing a reasonable balance between 
energetics and evolutionary timescales.

A broader comparison with previous CE and RLOF reconstructions reinforces this 
evolutionary pathway. Population studies generally recover mean efficiencies of 
$\alpha_{\rm CE}\!\approx\!0.3$–0.5 but with substantial system–to–system 
dispersion and no evidence for a universal value 
\citep[e.g.][]{2022MNRAS.513.3587Z,chenreview,santi2025}. 
\citet{2022MNRAS.513.3587Z} found moderate efficiencies in WD+BD systems, 
consistent with expectations for low-mass RGB donors, while 
\citet{chenreview} review a wider set of post-CE binaries—including double WDs— 
showing that $\alpha_{\rm CE}$ depends sensitively on donor structure, 
mass ratio, and boundary definition. In J2102--4145, the relatively high $\alpha_{\rm CE}$ follows directly 
from the very deep mass cut needed to leave the secondary with an 
extremely thin H envelope. Reproducing $M_{\rm H}\!\sim\!10^{-7}\,M_\odot$ 
requires removing tightly bound inner layers of the RGB envelope, which 
raises $E_{\rm bind}$ and thus $\alpha_{\rm CE}$. The large efficiency 
value therefore reflects the progenitor’s structure, not an anomalous CE 
process. We note that the orbital periods we infer at CE onset
($\sim$20--40\,d; Table~\ref{tab:master_CE}) agree well with the results from the independent
reconstruction method of \citet{2022MNRAS.513.3587Z}, lending further support to the viability
of this formation pathway.

A direct comparison can also be made with the adiabatic mass–loss
reconstructions of \citet{ge2022,ge2024}, where the CE evolution is
followed along adiabatic sequences that track the donor’s structural
readjustment under rapid mass removal while enforcing both the energy
and Roche–lobe conditions. Applied mainly to sdB+WD and sdB+MS
binaries, this method yields typical efficiencies of
$\alpha_{\rm CE}\!\sim\!0.3$, with noticeable system–to–system scatter
and a dependence on donor mass and mass ratio rather than a universal
value. In these models, the total H mass is integrated from the
$X_{\rm H}=0.1$ boundary outward, giving an upper limit immediately
after envelope ejection; the remnant is expected to lose additional
surface H as it thermally readjusts. According to updated
adiabatic grids provided by Hongwei~Ge (priv.~comm.), the residual
H masses predicted for the parameters of J2102--4145 range
between $M_{\rm H}\!\sim\!3\times10^{-4}$ and
$3\times10^{-3}\,M_\odot$ at the final orbital
separation, depending on the progenitor mass
and adopted efficiency. In contrast, reproducing the observed
radius of the secondary WD in J2102--4145 requires an extremely thin
residual envelope ($M_{\rm H}^{\rm fit}\!\simeq\!10^{-7}\,M_\odot$),
implying a deeper mass cut and higher $E_{\rm bind}$ than in the sdB
progenitors considered by \citet{ge2022,ge2024}, see later in this section. The larger
$\alpha_{\rm CE}$ values inferred here are therefore consistent with
their overall trends and reflect distinct structural conditions and
boundary definitions rather than fundamentally different CE physics.

To quantify the additional energy required to reach a given target efficiency, 
we parameterize the contribution of possible extra reservoirs (e.g. partial 
recombination or enthalpy) as a fraction \(f\) of the reference binding energy, 
such that \(E_{\rm bind}^{\rm eff}=E_{\rm bind}(1-f)\) and 
\(\alpha_{\rm CE}^{\rm new}=\alpha_{\rm CE}(1-f)\). The fraction needed to 
reach a target efficiency \(\alpha_\star\) then follows from 
\(f(\alpha_\star)=1-\alpha_\star/\alpha_{\rm CE}\), corresponding to an extra 
energy \(E_{\rm extra}=f\,E_{\rm bind}\).
The corresponding \(f\) and \(E_{\rm extra}\) values listed in 
Table~\ref{tab:master_CE} indicate that the \(1.5\,M_\odot\) progenitor 
requires only a negligible fraction (\(f\simeq0.003\)) to achieve 
\(\alpha_{\rm CE}\simeq1\), whereas reducing the efficiency to 
\(\alpha_\star=0.5\) would demand about 50\% of \(E_{\rm bind}\). For 
the \(1.8\,M_\odot\) model, the required fractions increase to 
\(f\simeq0.3\)–0.7 for the same range of \(\alpha_\star\). In contrast, 
the \(1.0\,M_\odot\) case already yields \(\alpha_{\rm CE}<1\) without 
extra energy and needs only \(\sim0.14\,E_{\rm bind}\) to reach 
\(\alpha_\star=0.5\).

The potential role of recombination energy can be assessed by estimating the 
maximum energy reservoir available if the envelope material above the cut were 
to recombine completely. Assuming a representative composition 
(\(X\simeq0.70\), \(Y\simeq0.28\)) and full recombination of H and He, the 
energy per unit mass is 
\begin{equation}
\epsilon_{\rm rec}\;\simeq\;
X\,\frac{13.6~\mathrm{eV}}{m_{\rm p}}
\;+\;Y\,\frac{79~\mathrm{eV}}{4\,m_{\rm p}}
\;\approx\;1.4\times10^{13}\ \mathrm{erg\,g^{-1}},
\label{eq:epsrec}
\end{equation}
where \(m_{\rm p}\) is the proton mass and \(1\,\mathrm{eV}=1.602\times10^{-12}\,\mathrm{erg}\).
The corresponding upper limit to the total 
recombination energy is then \(E_{\rm rec,max}\approx\epsilon_{\rm rec}M_{\rm env}
\approx2.9\times10^{46}(M_{\rm env}/M_\odot)\,\mathrm{erg}\), where 
\(M_{\rm env}=M_{\rm ZAMS}-m_{\rm cut}(M_{\rm H}^{\rm fit})\)\ \citep[e.g.][]{2016MNRAS.462..362I}.
For the three progenitors 
considered here, this yields \(E_{\rm rec,max}\simeq1.9\times10^{46}\), 
\(3.4\times10^{46}\), and \(4.3\times10^{46}\,\mathrm{erg}\) for 
\(M_{\rm ZAMS}=1.0\), 1.5, and 1.8\,$M_\odot$, respectively.
Comparing these values with the extra energies \(E_{\rm extra}\) in 
Table~\ref{tab:master_CE} shows that recombination could, in principle, 
supply enough energy only in the most favourable case 
(\(1.0\,M_\odot\), \(\alpha_\star=0.5\), 
\(E_{\rm extra}\simeq1.9\times10^{46}\) erg 
\(\approx E_{\rm rec,max}\)).
For the other cases, the energy required to reach the target efficiencies
(\(\alpha_\star=1\) or 0.5) exceeds \(E_{\rm rec,max}\) by factors of
$\sim$3–6, even under idealized assumptions of complete recombination and 
perfect energy trapping. In realistic CE conditions, partial recombination,
energy leakage, and compositional gradients would further reduce the effective
contribution. Therefore, except for the lowest-mass progenitor, recombination
alone is unlikely to provide the additional energy needed to lower
\(\alpha_{\rm CE}\) without violating the constraint on the thin H envelope.

\begin{figure}
  \centering
   \includegraphics[width=1.\columnwidth]{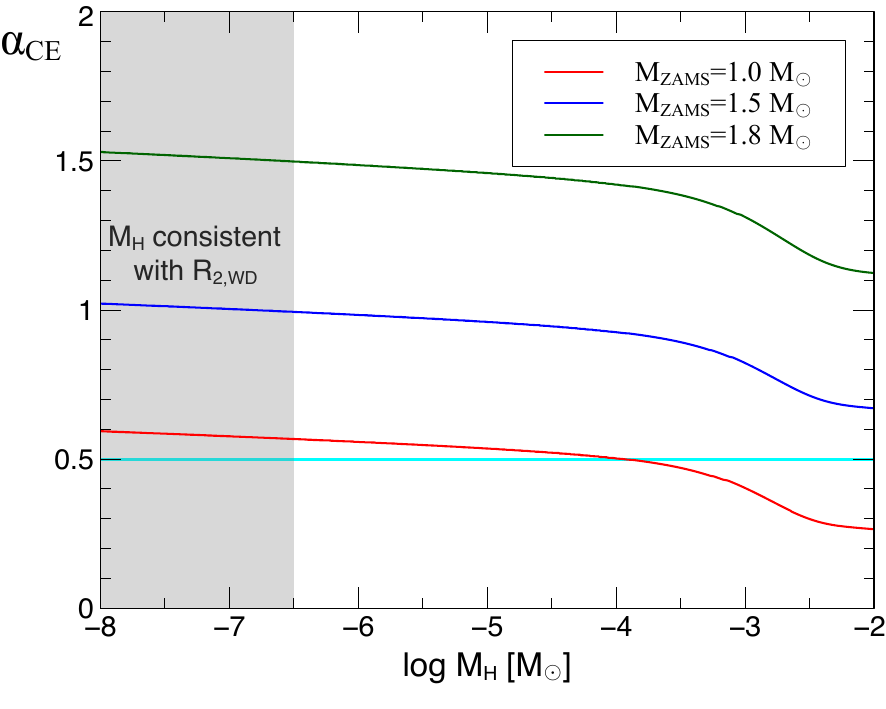}
  \caption{CE efficiency as a function of the residual H mass in the secondary WD.
    The grey band marks the range compatible with the observed radius of the
    secondary, \(\log M_{\rm H}\le -6.5\).}
  \label{alpha_vs_MH}
\end{figure}

Figure~\ref{alpha_vs_MH} illustrates the dependence of the CE efficiency 
on the residual H mass in the secondary WD. For all three progenitor 
models, \(\alpha_{\rm CE}\) flattens into a nearly constant plateau for 
\(\log M_{\rm H}\lesssim -6.5\), corresponding to the regime constrained by 
the observed radius of the secondary (grey band). In this range, the values 
of \(\alpha_{\rm CE}\) coincide with those listed in 
Table~\ref{tab:master_CE}, showing that the derived efficiencies are 
essentially insensitive to the exact choice of the residual H mass as long 
as it remains within the observationally allowed interval. 
At larger \(M_{\rm H}\), \(\alpha_{\rm CE}\) decreases monotonically because 
the binding energy becomes smaller when the envelope cut is placed further 
out in mass.

\begin{table*}
\centering
\caption{CE summary for J2102--4145, assuming a residual H mass 
\(M_{\rm H}^{\rm fit}=10^{-7}\,M_\odot\) that reproduces the observed radius 
of the secondary WD.}
\label{tab:master_CE}
\begingroup
\setlength{\tabcolsep}{4pt}
\small
\begin{tabular}{lcccccccccl}
\hline\hline
$M_{\rm ZAMS}$ & $M_{\rm core}$ & $R_{\rm phot}$ & $a_{\rm i}$ & $P_{\rm i}$ & $a_{\rm f}$ &
$\Delta E_{\rm orb}$ & $E_{\rm bind}$ & $\alpha_{\rm CE}$ &
$E_{\rm extra}(\alpha_\star)$ \\
$[M_\odot]$ & $[M_\odot]$ & $[R_\odot]$ & $[R_\odot]$ & [day] & $[R_\odot]$ &
$[10^{47}\,\mathrm{erg}]$ & $[10^{47}\,\mathrm{erg}]$ & &
$[10^{47}\,\mathrm{erg}]$ \\
\hline
1.0 & 0.315 & 28.0 & 60.1 & 37.0 & 0.854 & 2.494 & 1.44  & 0.577 &
$\,\{\alpha_\star{=}1:\ \text{---};\ \alpha_\star{=}0.5:\ 0.192\}$ \\
1.5 & 0.315 & 25.1 & 50.1 & 26.6 & 0.854 & 2.402 & 2.41  & 1.003 &
$\,\{\alpha_\star{=}1:\ 0.007;\ \alpha_\star{=}0.5:\ 1.21\}$ \\
1.8 & 0.315 & 22.1 & 42.7 & 20.6 & 0.854 & 2.315 & 3.494 & 1.510 &
$\,\{\alpha_\star{=}1:\ 1.18;\ \alpha_\star{=}0.5:\ 2.34\}$ \\
\hline
\end{tabular}
\par\endgroup

\vspace{0.25em}
\raggedright\footnotesize\textit{Notes:} 
\(a_{\rm now}=0.80190\,R_\odot\) (common to all three cases). 
\(P_{\rm i}\) is the orbital period at CE onset, derived from \(a_{\rm i}\) through Kepler’s third law 
using \(M_{1,{\rm WD}}=0.375\,M_\odot\). 
\(E_{\rm extra}(\alpha_\star)\) denotes the additional energy required to reach the target 
\(\alpha_\star\) while keeping \(M_{\rm H}^{\rm fit}=10^{-7}\,M_\odot\). 
The orbital-energy change is computed from \(a_{\rm i}\) to \(a_{\rm f}\), and the binding energy 
from \(m_{\rm cut}\) to the stellar surface.
\end{table*}

The stability criteria and RLOF reconstructions of \citet{ge2024}
provide a useful framework to contextualize the formation of the
primary WD. Figure~4 of \citet{ge2024} shows that stable, largely
non--conservative mass transfer on the RGB naturally produces He--core
remnants with $M_{\rm WD}\simeq0.30\,M_\odot$ for binaries with initial
orbital periods of $\sim$60\,d. Because the $P$--$M_{\rm WD}$ relation
in this mass range is set mainly by the core mass--radius relation of
the RGB donor and depends only weakly on $M_{\rm ZAMS}$, slightly
longer initial periods are expected to yield correspondingly more
massive remnants. In this context, the $0.375\,M_\odot$ primary in
J2102--4145 can be naturally interpreted as the outcome of stable RGB
mass transfer in a system with an initial period moderately longer
than the fiducial $\sim$60\,d case illustrated by \citet{ge2024}.

The observed $\sim$0.2\,Gyr cooling--age difference between the two WDs
is also compatible with this channel. In particular, the post--RLOF
orbital separation must allow the secondary to reach a CE phase within
the required timescale, which disfavors both fully conservative
transfer—leading to excessively wide orbits—and fully non--conservative
extremes, which would tend to produce lighter remnants or reduce the
likelihood of CE survival. While a detailed quantitative reconstruction
is beyond the scope of this work, these considerations indicate that
the first mass--transfer episode cannot lie at either extreme and is
likely to have been neither fully conservative nor fully
non--conservative.

\section{Conclusions}
\label{conclusions}

The eclipsing double WD system J2102--4145 provides a unique empirical
test of H-envelope retention and the outcome of envelope ejection in
low-mass CE events. From the observed parameters, both components are
unambiguously He-core WDs. C/O cores are excluded: the primary’s radius
cannot be reproduced by any C/O-core model at the observed $T_{\rm
eff}$, and the secondary’s mass ($0.314\,M_\odot$) is below the helium
ignition threshold.

Our analysis reveals a clear structural dichotomy. The less massive
secondary ($0.314\,M_\odot$) has a slightly smaller radius than the
primary ($0.375\,M_\odot$), a reversal of the expected mass–radius
trend that cannot be explained by the modest $\sim$800\,K temperature
difference. This contrast is explained by the primary retaining a
substantial H envelope, while the secondary has undergone nearly
complete stripping. Reproducing the latter’s radius requires
$M_{\rm H}\lesssim 10^{-7}\,M_\odot$, far below the values predicted by
standard bifurcation criteria. This difference shows that envelope removal
in J2102--4145 penetrated
much more deeply than predicted by standard prescriptions.
Rotational mixing could, in principle,
moderate the role of diffusion and affect the inferred envelope mass,
but the requirement of an extremely thin $M_{\rm H}$ remains robust.

We propose a formation sequence in which the primary formed first via
SRLOF, producing a He-core WD with a relatively large H envelope, and
the secondary subsequently formed through a CE episode that stripped
almost all of its envelope. Cooling ages support this scenario: the
secondary is matched by a thin-envelope CE track with a cooling age of
$\sim$220\,Myr, while the primary is consistent with SRLOF models that
retain thick H layers, yielding ages of $\sim$260\,Myr in our reference
case and up to $\sim$510\,Myr in an illustrative sequence with enhanced
residual burning ($L_{\rm nuc}/L\simeq0.70$). Even larger nuclear
contributions would further lengthen the cooling time. The resulting
age gap, of order a few $10^8$\,yr (within $\sim$40–290\,Myr in our
bracketing models), agrees with the proposed sequence where SRLOF
precedes CE. Independent constraints on the formation geometry indicate that the
progenitor of the less massive WD had a mass of $\sim$1.1–1.9\,
$M_\odot$ and an orbital period of $\sim$20–60\,d at CE onset,
consistent with the proposed SRLOF\,$\rightarrow$\,CE formation pathway.

Our CE energy analysis clarifies why the event that formed the secondary
must have removed the envelope to unusually deep layers.  The extremely
small residual H mass required by the observed radius
($M_{\rm H}\simeq 10^{-7}\,M_\odot$) forces a mass cut far below the usual
bifurcation criteria, resulting in a strongly bound remnant envelope and
the correspondingly large $\alpha_{\rm CE}$ values obtained in our models.
In contrast, adiabatic mass–loss reconstructions predict post–CE H layers
that are orders of magnitude larger, underscoring that J2102--4145 probes an
exceptionally stripped regime of low–mass CE evolution.  Nevertheless, this
extreme outcome fits naturally within the proposed SRLOF\,$\rightarrow$\,CE
formation sequence once the observational constraint on $M_{\rm H}$ is imposed.

J2102--4145 thus constitutes a benchmark for CE physics in the low-mass white-dwarf regime,
providing direct evidence of highly efficient envelope stripping
and placing stringent limits on post-CE H retention. Its precisely
measured parameters offer one of the strongest observational challenges
to standard prescriptions, and future discoveries of similar systems
will be essential to establish whether such extreme cases are unusual or
common outcomes of compact-binary evolution.

\begin{acknowledgements}
We thank the anonymous referee for a constructive and insightful
report that helped improve this manuscript.  We are also deeply
grateful to Hongwei~Ge for his kindness in sharing details of his
adiabatic mass–loss calculations and providing updated results for
direct comparison with our models. We are grateful to Yossef Zenati
for kindly providing helpful input during the preparation of this
work. We also thank Helena Mainetti for her help in preparing the
figures.  MZ and MV acknowledge support from FONDECYT (grant
1250525). LGA, ARM and ST acknowledge support from MINECO under the
PID2023-148661NB-I00 grant and by the AGAUR/Generalitat de Catalunya
grant SGR-386/2021.  This research has made use of NASA Astrophysics
Data System.
  
\end{acknowledgements}

\bibliographystyle{aa}
\bibliography{biblio}

\end{document}